\documentclass[10pt,conference]{IEEEtran}
\IEEEoverridecommandlockouts
\usepackage{cite}
\usepackage{subfig}
\usepackage{multirow}
\usepackage{graphicx}
\usepackage{arydshln}
\usepackage{pifont}
\usepackage{tcolorbox}
\usepackage{booktabs}
\usepackage{url}

\def\BibTeX{{\rm B\kern-.05em{\sc i\kern-.025em b}\kern-.08em
    T\kern-.1667em\lower.7ex\hbox{E}\kern-.125emX}}
\begin{document}

\title{Log Parsing using LLMs with Self-Generated In-Context Learning and Self-Correction}

\author{
\IEEEauthorblockN{
Yifan Wu,
Siyu Yu,
Ying Li
}
\IEEEauthorblockA{Peking University, Beijing, China}
\IEEEauthorblockA{\{yifanwu, li.ying\}@pku.edu.cn, gaiusyu6@gmail.com}

\thanks{Ying Li is the corresponding author.}}

\maketitle

\begin{abstract}
Log parsing transforms log messages into structured formats, serving as a crucial step for log analysis. Despite a variety of log parsers that have been proposed, their performance on evolving log data remains unsatisfactory due to reliance on human-crafted rules or learning-based models with limited training data. The recent emergence of large language models (LLMs) has demonstrated strong abilities in understanding natural language and code, making it promising to apply LLMs for log parsing. Consequently, several studies have proposed LLM-based log parsers. 
However, LLMs may produce inaccurate templates, and existing LLM-based log parsers directly use the template generated by the LLM as the parsing result, hindering the accuracy of log parsing.
Furthermore, these log parsers depend heavily on historical log data as demonstrations, which poses challenges in maintaining accuracy when dealing with scarce historical log data or evolving log data.
To address these challenges, we propose AdaParser, an effective and adaptive log parsing framework using LLMs with self-generated in-context learning (SG-ICL) and self-correction. To facilitate accurate log parsing, AdaParser incorporates a novel component, a template corrector, which utilizes the LLM to correct potential parsing errors in the templates it generates. 
In addition, AdaParser maintains a dynamic candidate set composed of previously generated templates as demonstrations to adapt evolving log data.
Extensive experiments on public large-scale datasets indicate that AdaParser outperforms state-of-the-art methods across all metrics, even in zero-shot scenarios. Moreover, when integrated with different LLMs, AdaParser consistently enhances the performance of the utilized LLMs by a large margin.
\end{abstract}


\section{Introduction}
Log messages are generated by logging statements in the source code to record system runtime information \cite{xu2024unilog,chen2021survey,chen2020studying}. By inspecting system logs, engineers can detect anomalies \cite{du2017deeplog}, localize software bugs \cite{chen2021pathidea}, or troubleshoot problems \cite{jia2021logflash} in the system. Therefore, log analysis plays a crucial role in software maintenance. The first step of log analysis is log parsing \cite{he2021survey}, which extracts two parts of log messages: (1) \textit{log templates}. i.e., constant parts that are explicitly written in logging statements, and (2) \textit{variables}, i.e., dynamic parts that vary during runtime and reflect system runtime status.

Manual log parsing is impractical due to the sheer volume of logs. Hence, numerous automatic log parsers have been proposed \cite{zhang2023system}, which can be categorized into syntax-based and semantic-based log parsers. 
Syntax-based log parsers \cite{fu2009execution,nagappan2010abstracting,he2017drain,dai2020logram} use human-crafted features or heuristics (e.g., log length and frequency) to extract log templates. 
In contrast, semantic-based log parsers \cite{liu2022uniparser, le2023alog, huo2023semparser, li2023did} employ deep learning models to mine semantics from log data for parsing.
However, recent studies have revealed that the performance of existing log parsers declines over time when handling evolving log data \cite{wang2022spine,yu2023log,liu2023logprompt}. 
On the one hand, syntax-based log parsers, which rely heavily on predefined rules, experience significant performance degradation when processing logs that deviate from these rules.
On the other hand, semantic-based log parsers, which typically employ deep learning models, struggle to adapt to evolving log data without periodic retraining, which is constrained by the quantity and quality of labeled data available.

To address these limitations, we intend to leverage the large language models (LLMs) to achieve adaptive log parsing for evolving log data. LLMs are trained on extensive corpora and exhibit strong capabilities in understanding natural language and code \cite{zhao2023survey, fan2023large}, making them well-suited for understanding log messages. Furthermore, LLM's powerful in-context learning (ICL) ability allows them to perform downstream tasks without model tuning \cite{dong2022survey, liu2023pre}. Although several studies \cite{liu2023logprompt,le2023blog,jiang2023lilac,xu2024divlog,ma2024llmparser} have explored using LLMs for log parsing, designing an effective and adaptive LLM-based log parser for evolving log data still faces the following challenges:

\textit{First}, despite their strong text understanding and generation abilities, LLMs may produce inaccurate parsing results, and existing LLM-based log parsers directly use the template generated by the LLM as the parsing result, hindering the accuracy of log parsing.
For example, the template generated by the LLM fails to accurately match the input log message or is overly broad, i.e., identifying crucial information as variables, which is essential for troubleshooting.

\textit{Second}, existing LLM-based parsers \cite{jiang2023lilac,xu2024divlog} heavily rely on historical log data to sample candidates as demonstrations. However, software updates often result in evolving log data, making sufficient historical log data unavailable. Our experiments indicate that these parsers perform well with abundant historical log data but struggle to maintain accuracy when historical log data is limited.

To tackle these challenges, we propose AdaParser, an effective and adaptive log parsing framework using LLMs with self-generated ICL (SG-ICL) and self-correction.
For each log message to be parsed, AdaParser initially employs a tree-based parser to determine if a corresponding template is already stored in the parsing tree. If so, AdaParser directly uses the matched template as the parsed result, thereby avoiding redundant LLM queries and enhancing parsing efficiency.
Otherwise, AdaParser leverages the LLM to parse logs. To mitigate performance degradation caused by evolving log data, we propose an SG-ICL-based parser. This parser maintains a dynamic candidate set composed of templates previously generated by the LLM and selects demonstrations from this candidate set to construct a prompt, guiding the LLM in parsing logs. 
To further reduce potential parsing errors in the templates, we propose a novel and effective template corrector that employs the LLM to correct parsing errors in the generated templates. The corrected templates are then used to update both the candidate set and the parsing tree.

We have conducted extensive experiments to evaluate AdaParser on public large-scale log datasets \cite{jiang2023large}.
The results show that AdaParser achieves a 96.5\% F1 score in grouping accuracy and an 87.9\% F1 score in template accuracy, outperforming state-of-the-art baselines and exhibiting remarkable robustness across diverse log datasets.
Furthermore, AdaParser surpasses all existing baselines in accuracy across various ratios of available logs, even in the absence of historical logs, underscoring its adaptability.
When integrated with different LLMs, AdaParser consistently enhances its performance on all metrics by a large margin.
In terms of efficiency, AdaParser achieves a speed comparable to the most efficient baseline, Drain \cite{he2017drain}.
The evaluation results demonstrate that AdaParser is an effective, adaptive, and efficient log parsing framework suitable for deployment in real-world evolving systems.

In summary, this paper makes the following contributions:
\begin{itemize}
    \item We propose AdaParser, an effective and adaptive LLM-based log parsing framework. With the proposed self-generated ICL and self-correction, AdaParser can perform accurate log parsing and adapt to evolving log data.
    \item We evaluate AdaParser on public large-scale datasets. The results show that AdaParser outperforms state-of-the-art methods on all metrics, even in zero-shot scenarios. Additionally, when integrated with different LLMs, AdaParser consistently enhances its performance by a large margin.
    \item We release the code of AdaParser at \url{https://github.com/wuyifan18/AdaParser} to benefit both practitioners and researchers in the field of log analysis.
\end{itemize}

\begin{figure}
    \centering
    \includegraphics[width=0.95\columnwidth]{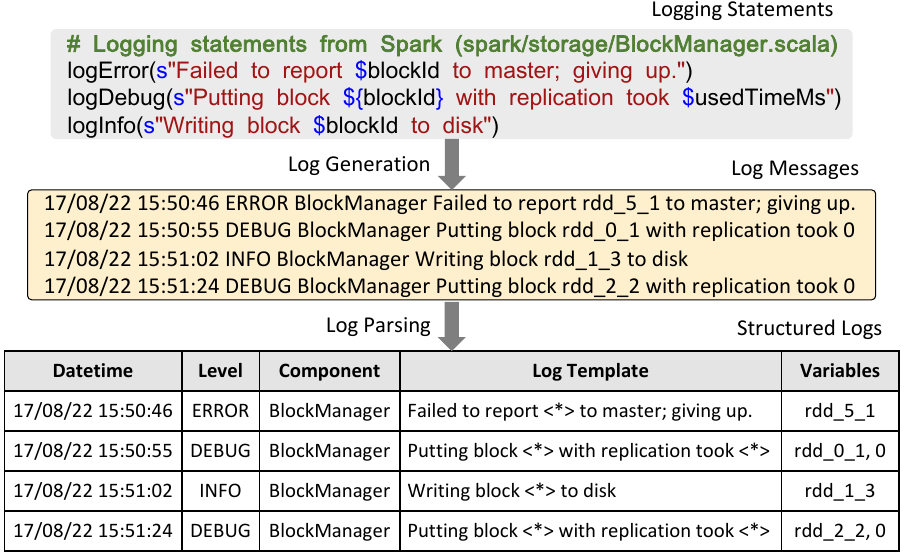}
    \caption{An example of log parsing from Spark.}
    \label{fig:example}
\end{figure}

\section{background and motivation}
\subsection{Log Parsing}
Log parsing is a fundamental step in log analysis \cite{he2021survey,zhang2023system}, converting semi-structured log messages to structured data. As illustrated in \figurename~\ref{fig:example}, in log parsing, a log parser is used to extract log headers (e.g., datetime ``\texttt{17/08/22 15:51:02}'', verbosity level ``\texttt{INFO}'', and component name ``\texttt{BlockManager}'') and log contents (e.g., ``\texttt{BlockManager Writing block rdd\_1\_3 to disk}''). 
Since log headers are automatically generated by the logger, their format is fixed and can usually be extracted easily. Therefore, log parsers primarily focus on extracting the constant parts (i.e., log templates) and the dynamic parts (i.e., variables) from log contents.
A straightforward method is to match log messages with corresponding logging statements within the source code \cite{pecchia2015industry,schipper2019tracing}. However, this method is impractical when the source code is inaccessible, such as commercial software.
Consequently, a variety of data-driven log parsers without requiring access to the source code have been proposed \cite{he2017drain,le2023alog,yu2023self}, which can be categorized into syntax-based and semantic-based methods.
Syntax-based log parsers \cite{fu2009execution,nagappan2010abstracting,he2017drain,dai2020logram} use human-crafted features or heuristics (e.g., log length and frequency) to extract log templates. 
In contrast, semantic-based log parsers \cite{liu2022uniparser, le2023alog, huo2023semparser, li2023did} employ deep learning models to mine semantics from log data for parsing.

In modern software systems, software updates lead to continuous changes in logging statements, resulting in evolving log data \cite{zhang2019robust,liu2023scalable,huo2023evlog}. For instance, Google has reported thousands of newly added logging statements each month due to software updates \cite{xu2010system}. Recent studies have indicated that existing log parsers' performance degrades over time when handling evolving log data \cite{wang2022spine,yu2023log,liu2023logprompt}. 
On the one hand, syntax-based log parsers heavily rely on human-crafted rules, requiring a substantial amount of domain-specific knowledge. This limitation results in poor performance when processing evolving log data that do not conform to these rules. For example, Drain \cite{he2017drain}, a leading syntax-based log parser, assumes that logs from the same template have the same length, which can lead to parsing errors when logs belonging to the same template vary in length.
On the other hand, semantic-based log parsers, which typically adopt deep learning models, struggle to adapt to evolving log data without periodic retraining, which is inherently limited by the quantity and quality of labeled data available for model tuning and involves expensive tuning processes. 
Therefore, there is a need for a log parser that does not rely on human-crafted rules and can adapt to evolving log data without requiring model tuning.
 
\subsection{Large Language Models}
Large Language Models (LLMs) are large-sized pre-trained language models with tens or hundreds of billions of parameters, trained on extensive unlabeled corpora via self-supervised learning. LLMs exhibit strong capabilities in understanding natural language \cite{zhao2023survey}. Beyond natural language, LLMs are also effective in understanding code, which has spurred interest in their application within the software engineering domain \cite{fan2023large}.
Effectively applying LLMs to downstream tasks is an important research topic. A common approach involves fine-tuning the model by learning the desired output from the given input of a downstream dataset to update model parameters \cite{wang2021codet5}.
However, fine-tuning can be extremely time-consuming and resource-intensive, especially for models with billions of parameters. Moreover, the performance of fine-tuned models heavily relies on the scale and quality of labeled data, making it less feasible in scenarios with limited labeled data.

Recently, \textit{in-context learning} (ICL) has emerged as a promising alternative that enables LLMs to perform downstream tasks without the necessity for explicit model tuning \cite{dong2022survey, liu2023pre}.
In the ICL paradigm, a formatted natural language prompt is used to query LLMs. This prompt typically includes three parts: (1) \textit{Instruction}, which describes the specific task; (2) \textit{Demonstrations}, which consist of several examples (i.e., query-answer pairs) selected from the labeled task datasets; and (3) \textit{Query}, which is the actual question that LLMs need to answer.
The number of demonstrations included in the prompt can vary, defining different scenarios. Specifically, if there are no demonstrations, the scenario is known as \textit{zero-shot learning}; if there is only one demonstration, the scenario is known as \textit{one-shot learning}; and \textit{few-shot learning} means there are several demonstrations. 
Numerous studies have demonstrated that ICL can significantly enhance the performance of LLMs across diverse tasks, such as log statement generation \cite{xu2024unilog}, comment generation \cite{geng2024large}, and code review \cite{guo2024exploring}. 
Given these advancements, we intend to apply the ICL paradigm to enable adaptive log parsing for evolving log data, capitalizing on the ability of LLMs to learn and adapt from structured examples without requiring continuous retraining.

\subsection{Challenges of Existing LLM-based parsers} \label{sec:challenges}
While recent advancements \cite{le2023blog,jiang2023lilac,liu2023logprompt,xu2024divlog,ma2024llmparser} have shown the potential of using LLMs for log parsing, these LLM-based parsers face the following critical challenges that limit their broader application.

\subsubsection{Unsatisfactory accuracy in parsing logs} 
\begin{figure}
    \centering
    \includegraphics[width=0.95\columnwidth]{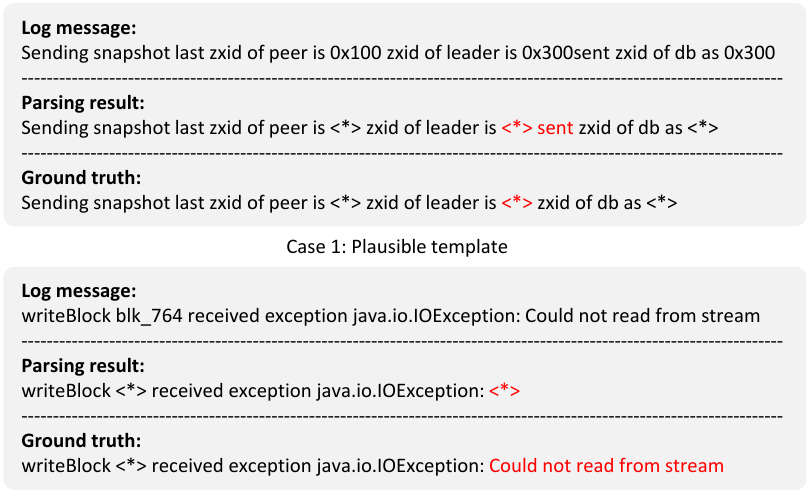}
    \caption{Two types of parsing errors in LLM-based parser.}
    \label{fig:cases}
\end{figure}

Despite their strong text understanding and generating capabilities, LLMs often produce inaccurate parsing results.
\figurename~\ref{fig:cases} illustrates two major types of parsing errors from LILAC \cite{jiang2023lilac}, a leading LLM-based parser: (1) \textit{Plausible template}, which occurs when the template generated by the LLM fails to accurately match the input log message (e.g., regular expression matching). Specifically, the LLM incorrectly parsed ``\texttt{0x300sent}'' as a variable ``\texttt{0x300}'' and a constant ``\texttt{sent}''. such errors reduce grouping accuracy and negatively impact downstream tasks like log compression and anomaly detection.
(2) \textit{Broad template}, which happens when the generated template matches the log message but is overly broad. Specifically, the LLM identified the exception message ``Could not read from stream'' as a variable. However, this exception message is crucial for troubleshooting and should be preserved in the log template. Such broad templates degrade template accuracy, affecting downstream tasks such as root cause analysis. Therefore, mitigating these parsing errors in the template generated by the LLM remains an unresolved challenge.

\subsubsection{Poor adaptability to evolving logs} Existing LLM-based parsers \cite{xu2024divlog, jiang2023lilac} rely heavily on historical log data to sample candidates as demonstrations. However, in practice, frequent software updates result in evolving logs, making sufficient historical log data unavailable. In extreme cases, such as when a brand-new service goes online, no historical log data exists. 
Our experiments also indicate that these parsers perform well with sufficient historical log data but struggle to maintain accuracy when such data is limited. For example, the template accuracy of LILAC decreases significantly by 31.2\% when in the absence of historical log data.
Consequently, enabling LLM-based log parsers to adapt to evolving logs without relying on extensive historical data is a substantial challenge.

In summary, while LLM-based log parsers show significant potential, several challenges must be addressed to enhance their accuracy and adaptability, which is essential to fully harness the potential of LLMs in log parsing and develop practical and effective solutions for real-world applications.

\begin{figure*}
    \centering
    \includegraphics[width=0.8\textwidth]{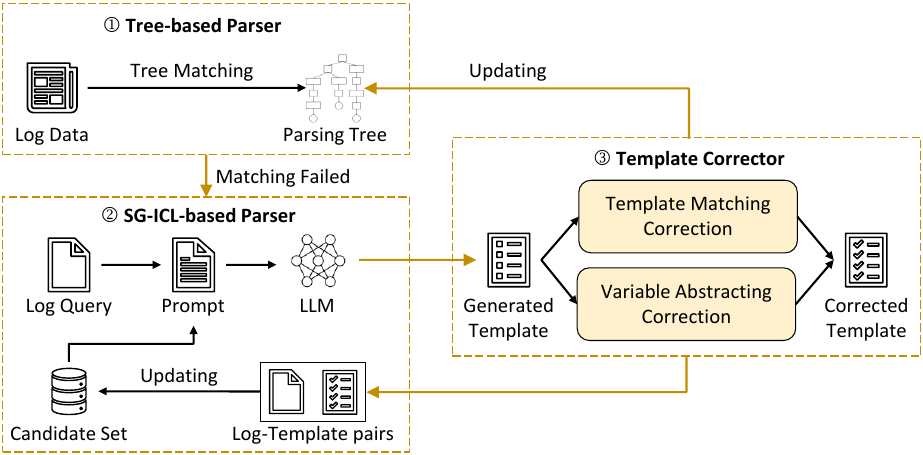}
    \caption{The overview of AdaParser framework.}
    \label{fig:overview}
\end{figure*}

\section{Approach}

\subsection{Overview}
To address the above challenges, we propose AdaParser, an LLM-based log parsing framework with self-generated ICL (SG-ICL) and self-correction, which can effectively parse logs and adapt to evolving log data. The overview of AdaParser is illustrated in \figurename~\ref{fig:overview}, which consists of three main components: \ding{172} \textit{Tree-based Parser}, \ding{173} \textit{SG-ICL-based Parser}, and \ding{174} \textit{Template Corrector}.

For each log message, AdaParser first employs the \textit{tree-based parser} to check whether its corresponding template is already stored in the parsing tree. If a match is found, AdaParser directly uses the matched template as the parsed result of this log message, which avoids redundant LLM queries and improves parsing efficiency. Otherwise, Adaparser utilizes the LLM to generate a new template for this log message. 
To mitigate the degradation of parsing accuracy due to evolving log data, we utilize the \textit{SG-ICL-based Parser} to maintain a dynamic candidate set using previously LLM-generated templates and select demonstrations from the candidate set to construct the parsing prompt, guiding the LLM to parse logs.
To reduce potential parsing errors in LLM-generated templates, we propose a novel and effective \textit{template corrector} to correct the template. The corrected template is then used to update both the candidate set and the parsing tree.

\subsection{Tree-based Log Parser}
In practical applications, the number of log templates is significantly smaller than the number of log messages \cite{jiang2023large,wang2022spine}. For instance, the Loghub-2.0 dataset \cite{jiang2023large} contains over 50 million logs but fewer than 3,500 templates. This disparity makes it inefficient to parse each log message individually using LLMs.
To improve efficiency, AdaParser employs a tree-based parser, using a trie-based parsing tree to store templates generated by the LLM and quickly match templates for new log messages. Each intermediate node in the tree represents a token from log templates, with the wildcard ``\verb|<*>|'' matching sequences of tokens of any length. The leaf node in the tree represents a complete log template, formed by concatenating all tokens along the path from the root to the leaf.

When a new log message arrives, AdaParser tokenizes it and matches the tokens against the parsing tree.
If the matching process reaches a leaf node, it indicates that the template stored at the leaf node exactly matches the log message, and AdaParser directly returns the template without further LLM querying.
If the matching process terminates at an intermediate node, the templates within the subtrees of the intermediate node form a list of relevant templates, denoted as $[T_1, T_2,..., T_n]$, which share the same prefixes as the input log message. AdaParser then utilizes the \textit{SG-ICL-based Parser} (Sec. \ref{sec:SG-ICL}) to parse the log message and generates a corrected template $T_c$ through the \textit{Template Corrector} (Sec. \ref{sec:self-correction}).
Subsequently, the parsing tree is updated based on the similarity between $T_c$ and the relevant templates.  
Specifically, we calculate the similarity between $T_c$ and each relevant template $T_i$ in $[T_1, T_2,..., T_n]$. We split them into lists of tokens, denoted as $L_c$ and $L_i$. The similarity between $L_c$ and $L_i$ is defined as follows:
\begin{equation}
\label{eq1}
  sim(T_c,T_i)=\frac{2 \times len(LCS(L_c, L_i))}{len(L_c)+len(L_i)}
\end{equation}
where LCS is the longest common subsequence of the two templates. 
If the similarity exceeds a predefined threshold (e.g., 0.8 in our implementation), it suggests that $T_c$ and $T_i$ are highly similar and likely derive from the same ground-truth template. AdaParser will merge $T_c$ and $T_i$ by replacing the different tokens in the path of $T_i$ with the wildcard ``\verb|<*>|''. 
If the similarity is below the threshold, it means that $T_c$ has a low correlation with $T_i$. However, templates with low similarity can derive from the same ground-truth template. For instance, in the Linux dataset from Loghub-2.0, the similarity between the log template ``apmd shutdown succeeded'' and relevant templates ``ntpd shutdown succeeded'', ``crond shutdown succeeded'', and ``xinetd shutdown succeeded'' is 0.67, which is below the threshold. Nonetheless, they all share the same ground-truth template ``\verb|<*>| shutdown succeeded''. To address this issue, we further calculate the number of different tokens in the same position among these templates with the same similarity. If the number exceeds a predefined threshold (e.g., 5 in our implementation), AdaParser merges $T_c$ and $T_i$. Otherwise, AdaParser inserts $T_c$ into the parsing tree. 

\begin{figure*}
    \centering
    \includegraphics[width=0.95\textwidth]{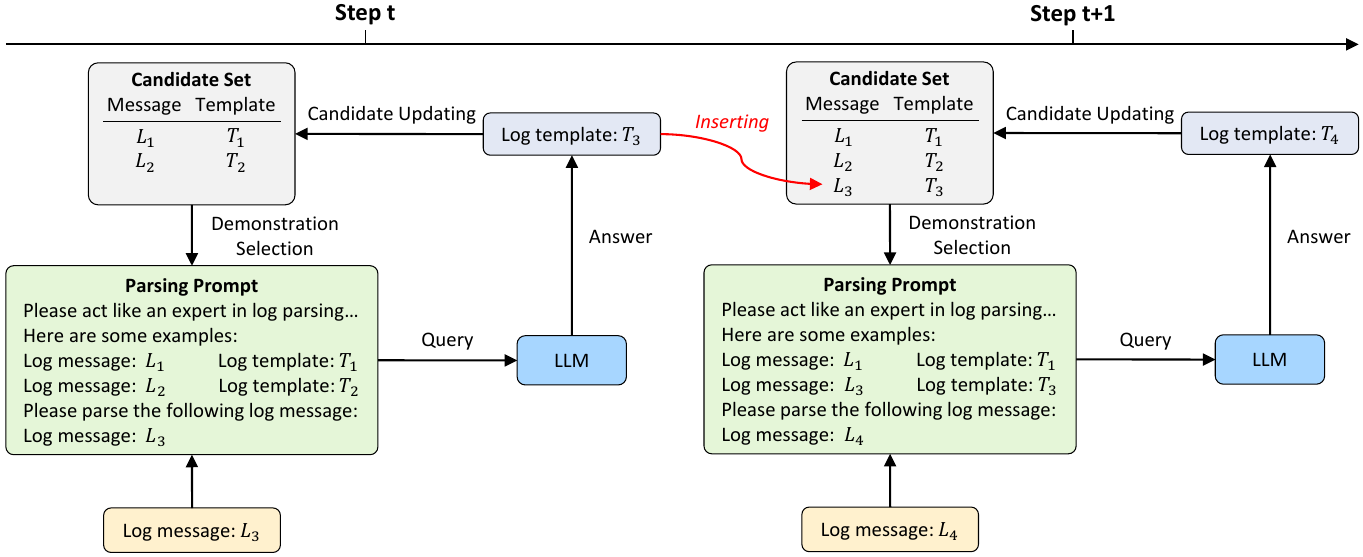}
    \caption{The workflow of SG-ICL-based parser. After step $t$, the message with the template generated by the LLM is added to the candidate set and then used as a demonstration at step $t+1$.}
    \label{fig:SG-ICL}
\end{figure*}

\subsection{SG-ICL-based Parser} \label{sec:SG-ICL}
The ICL paradigm initializes a candidate set by sampling a set of log messages and corresponding templates from the historical log data. LLMs can leverage system-specific features within demonstrations from the candidate set to facilitate more accurate log parsing. We utilize the hierarchical sampling algorithm \cite{jiang2023lilac} to extract a set of log messages from historical log data as the candidate set. However, as discussed in Sec. \ref{sec:challenges}, sufficient log data is not always available. Therefore, AdaParser supports initializing an \textit{empty} candidate set, which is incrementally populated by SG-ICL.

\figurename~\ref{fig:SG-ICL} illustrates the workflow of SG-ICL-based parser. At step $0$, if the candidate set is empty, we utilize zero-shot prompting to generate templates. At step $t$, when a new log message arrives, we first select similar demonstrations from the candidate set. 
Several studies have shown that demonstrations similar to the query can help LLMs better understand the desired behavior \cite{nashid2023retrieval, gao2023makes, geng2024large}.
Hence, we use a \emph{k}NN-based demonstration selection algorithm that is computationally efficient in practice. 
Specifically, for a queried log message $q$ and all candidate log messages $c_i$, we tokenized $q$ and $c_i$ into lists of tokens using delimiters such as punctuations and whitespace. Then, we calculate the LCS similarity $sim(q,c_i)$ as shown in Eq. \ref{eq1} and select the top-\emph{k} most similar logs from the candidate set as demonstrations. These logs, having similar tokens and special formats to the queried log, assist the LLM in understanding the semantics of the queried log.

These demonstrations are then assembled into the parsing prompt. Recent work \cite{gao2023makes} indicated that LLMs are prone to be influenced by the demonstrations that are closer to the query. Hence, we arrange the demonstrations in \textit{ascending order} of similarity to the queried log. This is based on the intuition that demonstrations with higher similarity may contain more information related to the queried log.
AdaParser then utilizes this parsing prompt to query the LLM and obtain the generated templates. Subsequently, a new example, i.e., the queried log message paired with the generated template, is added to the candidate set, facilitating continuous adaptation to evolving log data. At step $t+1$, the new example from the previous step $t$ becomes part of the candidate set and can be selected as a demonstration to help parse upcoming log messages.

\begin{figure*}
\centering
\subfloat[Template matching correction]{\includegraphics[width=0.95\columnwidth]{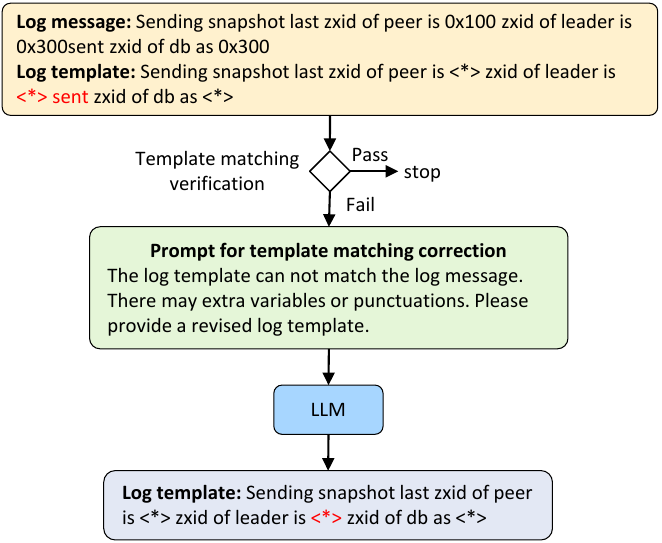}
\label{fig:template_matching}}
\hspace{1ex}
\subfloat[Variable abstracting correction]{\includegraphics[width=0.95\columnwidth]{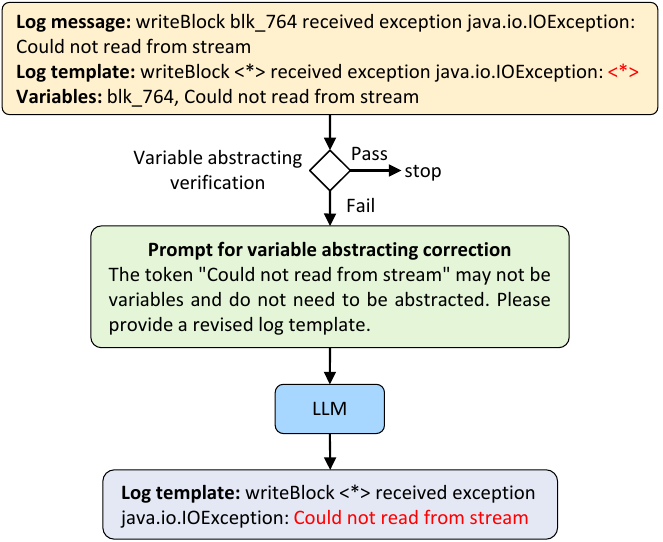}
\label{fig:variable_abstracting}}
\caption{Two examples of self-correction. Template matching correction and variable abstracting correction are used to correct the two types of parsing errors. The corrected parts are marked in red.}
\label{fig:self-correction}
\end{figure*}

\subsection{Template Corrector} \label{sec:self-correction}
Despite the considerable empirical success of LLMs, these models are not devoid of flaws. They occasionally exhibit undesired and inconsistent behaviors, such as generating seemingly convincing but inaccurate ``hallucinations'' and failing to trustfully adhere to rules and constraints \cite{pan2024automatically}. In particular, when applied to log parsing, LLMs are vulnerable to two types of parsing errors, as detailed in Sec. \ref{sec:challenges}. These errors can undermine parsing accuracy and pose obstacles for downstream tasks. 
A prevailing strategy to correct these undesired behaviors in LLMs involves learning from feedback, analogous to the typical human learning process where individuals refine their behaviors through cycles of trial, error, and correction. 
When making mistakes, humans often acquire feedback either from others or through self-reflection \cite{boyd1983reflective, metcalfe2017learning}. Such feedback provides valuable insights that enable humans to correct mistakes and modify their behavior accordingly. 

Inspired by this natural learning mechanism, we introduce a template corrector composed of two targeted self-correction strategies, i.e., template matching correction and variable abstracting correction, to correct the two primary types of parsing errors in LLM-based log parsers, respectively. The self-correction process iterates until the template either passes verification or a predefined number of iterations is reached. During each iteration, we gradually increase the temperature of the LLM to encourage more diverse and creative outputs. Specifically, the temperature at the $k$-th iteration is set to $k \times \alpha$, where $\alpha$ is a coefficient controlling the increase rate. Upon successful verification, the refined template is then used to update the candidate set and the parsing tree. Next, we illustrate the design details of these self-correction strategies.

\subsubsection{Template Matching Correction}
A correct template must accurately match the corresponding log message, otherwise, the template suffers from the error of \textit{Plausible template}. To fix this error, we propose a template matching correction strategy. Specifically, given a log message and a template generated by the LLM, we first verify whether the generated template exactly matches the input log message. This verification involves transforming the generated template into a regular expression (e.g., replace ``\verb|<*>|'' as ``.$*$'') to check if it matches the input log message exactly. If the regular expression fails to match the input log message, we employ a designed correcting prompt to rectify the template. As illustrated in \figurename~\ref{fig:template_matching}, the LLM initially parsed the token ``0x300sent'' incorrectly, treating it as a variable ``0x300'' and a constant ``sent''. This misinterpretation led to a mismatch between the input log message and the generated log template. After querying the LLM to revise the log template, the LLM recognized that “0x300sent” should be treated as a variable and abstracted it with the wildcard ``\verb|<*>|'', thereby successfully correcting the error in the initially generated template.

\subsubsection{Variable Abstracting Correction}
In log messages, certain key tokens (e.g., exception information) should not be abstracted as variables, otherwise, the log template exhibits the error of \textit{Broad template}. To fix this error, we propose a variable abstracting correction strategy. Specifically, given an input log message and a template generated by the LLM, we first extract the tokens that are replaced with the wildcard ``\verb|<*>|'' in the generated template. Next, we check whether these tokens contain or come after key tokens (e.g., ``Exception'', ``failed'', and ``interrupted'') that reflect system status and are critical for engineers. If so, we use a designed correcting prompt to correct the generated template. As illustrated in \figurename~\ref{fig:variable_abstracting}, the LLM initially abstracted the phrase ``Could not read from stream'' as a variable and replaced it with the wildcard ``\verb|<*>|''. However, because this phrase comes after the key token ``IOException'', it should not be abstracted. By prompting the LLM that the phrase may not be variables, the LLM correctly identified the phrase as constants, thereby correcting the error in the originally generated template.

\section{Experimental Setup}
\subsection{Research Questions}
We evaluate AdaParser on public large-scale log datasets by answering the following research questions:
\begin{itemize}
    \item RQ1: How effective is AdaParser?
    \item RQ2: Can AdaParser adapt to evolving logs?
    \item RQ3: How does each component contribute to AdaParser?
    \item RQ4: How generalizable is AdaParser with different LLMs?
    \item RQ5: How efficient is AdaParser?
\end{itemize}

\subsection{Datasets} \label{sec:datasets}
Our experiments are conducted using Loghub-2.0 \cite{jiang2023large}, a collection of large-scale annotated datasets for log parsing based on Loghub \cite{zhu2023loghub}. Loghub-2.0 includes ground-truth templates for 14 log datasets in Loghub sourced from various types of systems, such as distributed systems, supercomputers, operating systems, and server applications. On average, each dataset within Loghub-2.0 contains 3.6 million log messages, all labeled with ground-truth log templates. Additionally,  the total number of log templates across these datasets amounts to roughly 3,500.

\subsection{Evaluation Metrics}
Following recent studies \cite{khan2022guidelines,jiang2023large,jiang2023lilac}, we used the following four metrics in our experiments:

\subsubsection{Grouping Accuracy (GA)}
GA \cite{zhu2019tools} is a log-level metric that assesses the ability to correctly group log messages belonging to the same template.
It is defined as the ratio of correctly grouped log messages over all log messages. A log message is regarded as correctly grouped if and only if its template has the same group of log messages as the oracle.

\subsubsection{F1 score of Grouping Accuracy (FGA)}
FGA \cite{jiang2023large} is a template-level metric that measures the proportion of correctly grouped templates.
It is computed as the harmonic mean of precision and recall of grouping accuracy, where the template is considered as correct if log messages of the predicted template have the same group of log messages as the oracle.

\subsubsection{Parsing Accuracy (PA)}
PA is a log-level metrics \cite{dai2020logram} that measures the correctness of extracted templates and variables.
It is defined as the ratio of correctly parsed log messages to the total number of log messages. A log message is regarded as correctly parsed if and only if all tokens of templates and variables are identical with the oracle.

\subsubsection{F1 score of Template Accuracy (FTA)}
FTA is a template-level metric that measures the proportion of correctly identified templates. It is computed as the harmonic mean of precision and recall of Template Accuracy. A template is regarded as correctly identified if and only if log messages of the parsed template share the same ground-truth template and all tokens of the template are the same as those of the ground-truth template.

\subsection{Baselines}
In line with recent benchmark studies \cite{khan2022guidelines,jiang2023large}, we select five open-source and state-of-the-art log parsers for comparison with our method. The first two, AEL \cite{jiang2008abstracting} and Drain \cite{he2017drain}, are chosen due to their superior performance among all syntax-based log parsers. We also choose two current top-performing semantic-based log parsers, UniParser \cite{liu2022uniparser} and LogPPT \cite{le2023alog}. Furthermore, We incorporate the latest LLM-based log parser, LILAC \cite{jiang2023lilac}, which has shown the highest parsing
accuracy on Loghub-2.0. To ensure a fair comparison, we utilize the implementations of all baselines from their replication repositories, choosing the default settings or hyperparameters.

\subsection{Environment and Implementation}
We conducted experiments on a GPU Server equipped with an
NVIDIA A100 GPU since UniParser and LogPPT require GPU resources to perform log parsing. 
The default LLM in AdaParser is set to ChatGPT \cite{ChatGPT} (gpt-3.5-turbo-0125), primarily due to its popularity in
recent research \cite{le2023blog,xu2024divlog,jiang2023lilac}. 
ChatGPT was accessed via the OpenAI API, with both temperature and seed set to 0 to ensure the deterministic output for the same query, thereby enhancing reproducibility. Furthermore, we employ different LLMs to assess the generalizability of AdaParser in RQ4. 

To ensure fair evaluation, we adhere to the LLM-based baseline (i.e., LILAC) by default, utilizing the first 20\% of the logs from each dataset as available logs. Candidates are selected using the hierarchical candidate sampling algorithm, with the default number of candidates and demonstrations set to 32 and 3, respectively. Furthermore, we vary the available logs with different ratios to evaluate the adaptability of AdaParser in RQ2.
We have implemented AdaParser in Python and integrated it into previous benchmarks \cite{zhu2019tools,khan2022guidelines,jiang2023large}, allowing for a fair comparison between AdaParser and all baselines in the same framework. For all experiments that exhibit randomness, we repeat them five times and report their mean as the final result to mitigate potential random bias.

\begin{table*}
\centering
\caption{Accuracy comparison between AdaParser and baselines (\%).}
\label{tab:RQ1}
\resizebox{\textwidth}{!}{
\begin{tabular}{c|cccc|cccc|cccc|cccc|cccc|cccc}
\specialrule{0.4mm}{0pt}{0pt}
\multirow{2}{*}{Dataset} & \multicolumn{4}{c|}{AEL}                  & \multicolumn{4}{c|}{Drain}                & \multicolumn{4}{c|}{UniParser} & \multicolumn{4}{c|}{LogPPT} & \multicolumn{4}{c|}{LILAC}                 & \multicolumn{4}{c}{AdaParser}                                 \\ 
                         & GA           & FGA          & PA   & FTA  & GA           & FGA          & PA   & FTA  & GA     & FGA   & PA    & FTA   & GA    & FGA   & PA   & FTA  & GA            & FGA          & PA   & FTA  & GA            & FGA           & PA            & FTA           \\ \hline
Hadoop                   & 82.3         & 11.7         & 53.5 & 5.8  & 92.1         & 78.5         & 54.1 & 38.4 & 69.1   & 62.8  & 88.9  & 47.6  & 48.3  & 52.6  & 66.6 & 43.4 & 91.4          & 92.6         & 88.0 & 76.9 & \textbf{98.8} & \textbf{96.2} & \textbf{96.5} & \textbf{86.1} \\
HDFS                     & 99.9         & 76.4         & 62.1 & 56.2 & 99.9         & 93.5         & 62.1 & 60.9 & \textbf{100}    & 96.8  & 94.8  & 58.1  & 72.1  & 39.1  & 94.3 & 31.2 & \textbf{100}           & 96.8         & 99.9 & 92.5 & \textbf{100}  & \textbf{100}  & \textbf{100}  & \textbf{100}  \\
OpenStack                & 74.3         & 68.2         & 2.9  & 16.5 & 75.2         & 0.7          & 2.9  & 0.2  & \textbf{100}    & 96.9  & 51.6  & 28.9  & 53.4  & 87.4  & 40.6 & 73.8 & \textbf{100}  & \textbf{100} & 97.0 & 91.7 & \textbf{100}  & \textbf{100}  & \textbf{100}  & \textbf{97.9} \\
Spark                    & -            & -            & -    & -    & 88.8         & 86.1         & 39.4 & 41.2 & 85.4   & 2.0   & 79.5  & 1.2   & 47.6  & 37.4  & 95.2 & 29.9 & 88.3          & 82.5         & 88.9 & 71.3 & \textbf{98.5} & \textbf{94.1} & \textbf{98.4} & \textbf{84.0}   \\
Zookeeper                & 99.6         & 78.8         & 84.2 & 46.5 & 99.4         & 90.4         & 84.3 & 61.4 & 98.8   & 66.1  & \textbf{98.8}  & 51.0  & 96.7  & 91.8  & 84.5 & 80.9 & \textbf{99.7}          & 94.3         & 82.4 & 83.9 & 99.3 & \textbf{97.8} & 96.8 & \textbf{93.3} \\
BGL                      & 91.5         & 58.7         & 40.6 & 16.5 & 91.9         & 62.4         & 40.7 & 19.3 & 91.8   & 62.4  & 94.9  & 21.9  & 24.5  & 25.3  & 93.8 & 26.1 & 91.2          & 85.4         & 97.5 & 78.0 & \textbf{93.5} & \textbf{94.6} & \textbf{98.3} & \textbf{84.9} \\
HPC                      & 74.8         & 20.1         & 74.1 & 13.6 & 79.3         & 30.9         & 72.1 & 15.2 & 77.7   & 66.0  & 94.1  & 35.1  & 78.2  & 78.0  & 99.7 & 76.8 & 86.9          & 86.7         & 94.0 & 76.9 & \textbf{100}  & \textbf{100}  & \textbf{100}  & \textbf{95.9} \\
Thunderbird              & 78.6         & 11.6         & 16.3 & 3.5  & 83.1         & 23.7         & 21.6 & 7.1  & 57.9   & 68.2  & 65.4  & 29.0  & 56.4  & 21.6  & 40.1 & 11.7 & 79.2          & 86.6         & 52.7 & 56.0 & \textbf{91.1} & \textbf{91.8} & \textbf{72.6} & \textbf{68.0}   \\
Linux                    & 91.6         & 80.6         & 8.2  & 21.7 & 68.6         & 77.8         & 11.1 & 25.9 & 28.5   & 45.1  & 16.4  & 23.2  & 20.5  & 71.2  & 16.8 & 42.8 & 95.4          & 89.7         & 84.9 & 60.3 & \textbf{98.6} & \textbf{92.3} & \textbf{91.9} & \textbf{77.1} \\
Mac                      & 79.7         & 79.3         & 24.5 & 20.5 & 76.1         & 22.9         & 35.7 & 6.9  & 73.7   & 69.9  & 68.8  & 28.3  & 54.4  & 49.3  & 39.0 & 27.4 & 80.5          & 86.4         & 60.8 & 54.9 & \textbf{93.2} & \textbf{90.1} & \textbf{73.4} & \textbf{64.1} \\
Apache                   & \textbf{100} & \textbf{100} & 72.7 & 51.7 & \textbf{100} & \textbf{100} & 72.7 & 51.7 & 94.8   & 68.7  & 94.2  & 26.9  & 78.6  & 60.5  & 94.8 & 36.8 & \textbf{100}  & \textbf{100} & \textbf{99.9} & 89.7 & \textbf{100}  & \textbf{100}  & \textbf{99.9} & \textbf{93.1} \\
OpenSSH                  & 70.5         & 68.9         & 36.4 & 33.3 & 70.7         & 87.2         & 58.6 & 48.7 & 27.5   & 0.9   & 28.9  & 0.5   & 27.7  & 8.1   & 65.4 & 10.5 & 74.8          & 87.7         & 99.9 & 87.7 & \textbf{74.9} & \textbf{94.7} & \textbf{100}  & \textbf{94.7} \\
HealthApp                & 72.5         & 0.8          & 31.1 & 0.3  & 86.2         & 1.0          & 31.2 & 0.4  & 46.1   & 74.5  & 81.7  & 46.2  & 99.8  & 94.7  & \textbf{99.7} & 82.2 & 99.3 & 97.1         & 73.3 & 85.1 & \textbf{100}  & \textbf{100}  & 85.4 & \textbf{91.7} \\
Proxifier                & 97.4         & 66.7         & 67.7 & 41.7 & 69.2         & 20.6         & 68.8 & 17.6 & 50.9   & 28.6  & 63.4  & 45.7  & 98.9  & 87    & \textbf{100}  & 95.7 & \textbf{100}           & \textbf{100}          & \textbf{100}  & \textbf{100}  & \textbf{100}  & \textbf{100}  & \textbf{100}  & \textbf{100}  \\ \hline
Average                  & 85.6         & 55.5         & 44.2 & 25.2 & 84.3         & 55.4         & 46.8 & 28.2 & 71.6   & 57.8  & 73.0  & 31.7  & 61.2  & 57.4  & 73.6 & 47.8 & 91.9          & 91.8         & 87.1 & 78.9 & \textbf{96.3} & \textbf{96.5} & \textbf{93.8} & \textbf{87.9} \\ \specialrule{0.4mm}{0pt}{0pt}
\end{tabular}
}
\end{table*}

\begin{figure*}
    \centering
    \includegraphics[width=0.95\textwidth]{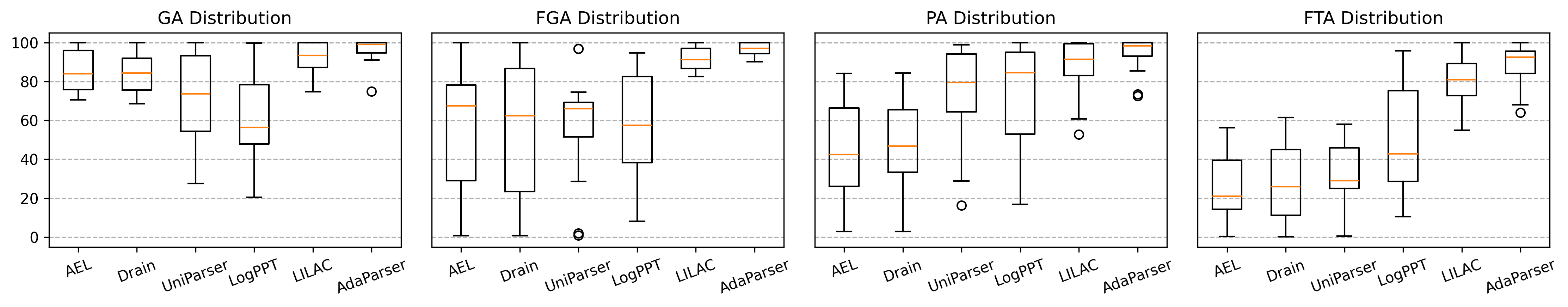}
    \caption{Robustness comparison between AdaParser and baselines (\%).}
    \label{fig:robustness}
\end{figure*}

\section{Evaluation Results}
\subsection{RQ1: How effective is AdaParser?}
In this RQ, we evaluate the performance of AdaParser in terms of accuracy and robustness, comparing it to other state-of-the-art baselines on public datasets.

\subsubsection{Accuracy}
Accuracy is the most critical indicator to evaluate the effectiveness of log parsers. We utilize the default settings of all methods and apply them to all log datasets. The results are shown in Table~\ref{tab:RQ1} with the best results for each metric on each dataset marked in \textbf{bold}. Metrics for AEL on the Spark dataset are denoted as “–” since it fails to complete the parsing process within a reasonable time (i.e., 12 hours), following previous works \cite{jiang2023large,jiang2023lilac}.

The results demonstrate that AdaParser outperforms all baselines on all average
metrics. Specifically, in terms of group-related metrics (i.e., GA and FGA), AdaParser achieves average scores of 96.3\% and 96.5\% on GA and FGA, outperforming the best syntax-based baseline, Drain, by 14.2\% in GA and 74.2\% in FGA. 
When considering the metrics related to parsing ability (i.e., PA and FTA), LILAC achieves the highest PA of 87.1\% and FTA of 78.9\% among all baselines. However, AdaParser achieves superior parsing metrics, with a PA of 93.8\% and an FTA of 87.9\%, which outperforms LILAC by 7.7\% and 11.4\%, respectively. Most notably, in terms of FTA, which is considered the most rigorous and comprehensive metric, AdaParser surpasses all baselines across all datasets. Given the strict definitions of correctly parsed and correctly identified, achieving such high metrics indicates that AdaParser possesses a strong capacity to distinguish between log templates and variables.

\subsubsection{Robustness}
Robustness is also an essential indicator for evaluating the effectiveness of log parsers. The strong robustness implies that log parsers can maintain a stable performance across log data with diverse characteristics, thereby demonstrating superior generalizability \cite{zhu2019tools,jiang2023large,le2023alog,xu2024divlog}.
To compare the robustness of AdaParser with all baselines, we illustrate the distribution of each log parser’s metrics across all datasets using box plots.

The results are shown in \figurename~\ref{fig:robustness}. AdaParser not only achieves the highest accuracy but also exhibits the smallest variance in performance, as evidenced by its narrowest distribution range. This demonstrates that AdaParser exhibits the strongest robustness in parsing a variety of log data. Specifically, the standard deviations of AdaParser for GA, FGA, PA, and FTA are 6.8\%, 3.6\%, 9.7\%, and 11.4\%, respectively, outperforming the currently most robust log parser, LILAC, which has standard deviations of 8.8\%, 6.2\%, 15.2\%, and 14.0\%. 
The strong robustness of AdaParser is primarily derived from the extensive pre-trained knowledge in LLMs related to diverse log data from different systems. In addition, the SG-ICL-based parser in AdaParser adapts the LLM to the system-specific characteristics of log datasets and is not subject to historical logs, which can often undermine robustness when dealing with diverse log data.

\begin{tcolorbox}
\textbf{Answer to RQ1:} AdaParser outperforms baselines in both message-level metrics and template-level metrics. Furthermore, AdaParser exhibits the strongest robustness when parsing diverse logs from different systems.
\end{tcolorbox}

\subsection{RQ2: Can AdaParser adapt to evolving logs?}
In this RQ, we investigate the adaptability of AdaParser to evolving logs and compare its performance to the current state-of-the-art LLM-based log parser, LILAC. To simulate log evolution, we vary the availability of historical logs, decreasing from 20\% (the default setting consistent with LILAC) to 0\% (representing a zero-shot scenario), with different template coverage ranging from 67.1\% to 0\%. For example, with 20\% of historical logs available, we assume that 32.9\% of the total templates are new log templates resulting from log evolution. Furthermore, in a zero-shot scenario, all templates are considered new for log parsers, simulating an extreme case, such as the launch of a brand-new service.

Table~\ref{tab:ratio} presents the experimental results. We observe that the performance of both AdaParser and LILAC decreases as the proportion of available logs decreases. Nevertheless, AdaParser consistently outperforms LILAC across all ratios of available logs. Remarkably, even in the zero-shot scenario, AdaParser achieves an accuracy that exceeds all existing baselines. Specifically, AdaParser outperforms LILAC by 29.1\% and 44.9\% for PA and FTA. This highlights the practicality of AdaParser in real-world evolving systems.
Furthermore, AdaParser demonstrates a smaller decline in performance compared to LILAC when faced with evolving logs.
Specifically, as the ratio of available logs decreases from 20\% to 0\%, LILAC's PA and FTA experience a considerable decrease of 28.5\% and 31.2\%, respectively. In contrast, AdaParser shows a more modest reduction of 14.3\% in PA and 10.5\% in FTA. These results emphasize the remarkable adaptability of AdaParser to evolving logs.

\begin{tcolorbox}
\textbf{Answer to RQ2:} AdaParser achieves superior performance compared to LLM-based parsers across various ratios of available logs. Even in the absence of historical logs, AdaParser achieves an accuracy surpassing all baselines, underscoring its remarkable adaptability.
\end{tcolorbox}

\begin{table}
\centering
\caption{Average accuracy among AdaParser with different ratios of available logs.}
\label{tab:ratio}
\resizebox{\columnwidth}{!}{
\begin{tabular}{c|c|l|cccc}
\specialrule{0.4mm}{0pt}{0pt}
Log Ratio & Template Coverage & Approach            & GA                   & FGA                  & PA                   & FTA                  \\ \hline
\multirow{4}{*}{20\%} & \multirow{4}{*}{2340/3488 (67.1\%)}  & LILAC               & 91.9                 & 91.8                 & 87.1                 & 78.9                 \\
                        & & AdaParser           & 96.3                 & 96.5                 & 93.8                 & 87.9                 \\ \cdashline{3-7}
                        & & w/o SG-ICL          & 95.5                 & 95.5                 & 93.2                 & 86.9                 \\
                        & & w/o self-correction & 95.4                 & 95.7                 & 91.8                 & 85.6                 \\  \hline
\multirow{4}{*}{1\%} & \multirow{4}{*}{1415/3488 (40.6\%)}   & LILAC               & 91.5                 & 92.0                 & 83.8                 & 78.1                 \\
                        & & AdaParser           & 95.5                 & 96.1                 & 92.2                 & 86.5                 \\
                        \cdashline{3-7}
                        & & w/o SG-ICL          & 95.2                 & 95.3                 & 92.0                 & 85.6                 \\
                        & & w/o self-correction & 95.2                 & 94.9                 & 91.6                 & 83.1                 \\ \hline
\multirow{4}{*}{0.1\%} & \multirow{4}{*}{876/3488 (25.1\%)} & LILAC    & 90.7                 & 91.0                 & 82.6     & 75.3                     \\
                        & & AdaParser           & 93.8                 & 94.2                 & 90.9                 & 84.2                 \\
                        \cdashline{3-7}
                        & & w/o SG-ICL          & 93.6                 & 92.7                 & 90.5                 & 83.1                 \\
                        & & w/o self-correction & 93.2                 & 93.0                 & 89.1                 & 79.2                 \\ \hline
\multirow{4}{*}{0.01\%} & \multirow{4}{*}{512/3488 (14.7\%)} & LILAC           & 89.0                 & 86.8                 & 77.1                 & 69.9                 \\
                        & & AdaParser           & 93.4                 & 93.8                 & 90.0                 & 83.6                 \\
                        \cdashline{3-7}
                        & & w/o SG-ICL          & 93.2                 & 91.9                 & 89.0                 & 79.8                 \\
                        & & w/o self-correction & 93.3                 & 90.7                 & 88.3                 & 77.1                 \\ \hline
\multirow{4}{*}{0\%} & \multirow{4}{*}{0/3488 (0\%)}   & LILAC               & 83.1                 & 76.2                 & 62.3                 & 54.3                 \\
                        & & AdaParser           & 86.1                 & 89.3                 & 80.4                 & 78.7                 \\
                        \cdashline{3-7}
                        & & w/o SG-ICL          & 85.5                 & 88.2                 & 77.5                 & 76.1                 \\
                        & & w/o self-correction & 82.6                     & 84.7                     & 73.5                     & 67.2                     \\ \specialrule{0.4mm}{0pt}{0pt}
\end{tabular}%
}
\end{table}

\subsection{RQ3: How does each design contribute to AdaParser?}
In this RQ, we conduct an ablation study to investigate the contributions of two designed components within AdaParser, i.e., the SG-ICL-based parser and the template corrector, under different ratios of available logs.
Specifically, we replaced SG-ICL with vanilla ICL that utilizes fixed candidate sets (denoted as w/o SG-ICL) and removed the template corrector, relying solely on the LLM's outputs as the final parsing results (denoted as w/o self-correction).

The results are shown in Table~\ref{tab:ratio}, in which the following observations can be made:
(1) Removing SG-ICL or self-correction leads to a decrease across all metrics under various ratios of available logs. 
For instance, removing self-correction leads to a decrease of 8.6\% and 14.6\% in PA and FTA under zero-shot scenarios, respectively. This decrease is primarily due to the parsing errors in the templates generated by the LLM, indicating that self-correction can mitigate parsing errors and enhance parsing accuracy.
(2) The contribution of our designed components becomes increasingly significant as the ratio of available logs decreases. This is because the LLM acquires task-specific knowledge from sufficient available logs through ICL, enabling it to directly generate correct templates. Conversely, when available logs are scarce or entirely absent, the LLM struggles to generate accurate templates. In these cases, our designed components are fully utilized to enhance parsing accuracy.
(3) Incorporating SG-ICL into AdaParser not only enhances accuracy but also efficiency. For example, in zero-shot scenarios, AdaParser with SG-ICL processes each dataset in an average of 673 seconds, compared to 839 seconds without SG-ICL.
Moreover, our statistical analysis reveals a substantial decrease in the number of self-corrections required for templates generated by the LLM when equipped with SG-ICL.
Specifically, the average number of self-correction per dataset for AdaParser with SG-ICL is 40.1, while that for AdaParser without SG-ICL is 83.7. Therefore, this improvement in efficiency can largely be attributed to SG-ICL's superior adaption to evolving logs, which reduces the number of generated incorrect templates and the consequent need for self-correction.
In conclusion, the ablation study demonstrates that both SG-ICL and self-correction are critical to AdaParser's performance, especially in scenarios with limited available logs.

\begin{tcolorbox}
\textbf{Answer to RQ3:} Both designs of SG-ICL and self-correction contribute to enhancing AdaParser's performance, particularly under limited log availability.
\end{tcolorbox}

\begin{table}
\centering
\caption{The performance of AdaParser with different LLMs under zero-shot scenario (\%).}
\label{tab:backbone}
\resizebox{\columnwidth}{!}{%
\begin{tabular}{c|c|cccc}
\specialrule{0.4mm}{0pt}{0pt}
Model                            & Approach  & GA         & FGA        & PA         & FTA        \\ \hline
\multirow{2}{*}{gpt-3.5-turbo}         & Base      & 81.2       & 85.0       & 70.2       & 68.8       \\
                                 & AdaParser & 86.1 ($\uparrow$ 6.0\%) & 89.3 ($\uparrow$ 5.1\%) & 80.4 ($\uparrow$ 14.5\%) & 78.7 ($\uparrow$ 14.4\%) \\ \hline
\multirow{2}{*}{gemini-1.5-flash}  & Base      & 79.8       & 80.3       & 61.3       & 55.2       \\
                                 & AdaParser & 83.2 ($\uparrow$ 4.3\%) & 84.2 ($\uparrow$ 4.9\%) & 77.6 ($\uparrow$ 26.6\%) & 69.8 ($\uparrow$ 26.4\%) \\ \hline
\multirow{2}{*}{claude-3-sonnet} & Base      & 78.4       & 79.9       & 63.7       & 56.9       \\
                                 & AdaParser & 84.5 ($\uparrow$ 7.8\%) & 87.6 ($\uparrow$ 9.6\%) & 78.5 ($\uparrow$ 23.2\%) & 73.4 ($\uparrow$ 29.0\%) \\ \hline
\multirow{2}{*}{DeepSeek-V2-chat}     & Base      & 79.4       & 77.0       & 63.2       & 56.7       \\
                                 & AdaParser & 81.8 ($\uparrow$ 3.0\%) & 84.3 ($\uparrow$ 9.5\%) & 72.7 ($\uparrow$ 15.0\%) & 68.9 ($\uparrow$ 21.5\%) \\ \hline
\multirow{2}{*}{qwen1.5-72b-chat}     & Base      & 75.9       & 70.0       & 60.6       & 50.0       \\
                                 & AdaParser & 79.4 ($\uparrow$ 4.6\%) & 76.7 ($\uparrow$ 9.6\%) & 72.8 ($\uparrow$ 20.1\%) & 63.4 ($\uparrow$ 26.8\%) \\ \specialrule{0.4mm}{0pt}{0pt}
\end{tabular}%
}
\end{table}

\subsection{RQ4: How generalizable is AdaParser with different LLMs?} \label{sec:RQ4}
In this RQ, we assess the performance of AdaParser with different LLMs under zero-shot scenarios, i.e., no historical data is available. Our goal is to determine if the performance gains from AdaParser can be generalized to other LLMs. We extend our evaluation beyond the default GPT-3.5 to include four additional representative and popular LLMs, including two closed-source LLMs (i.e., gemini-1.5-flash \cite{reid2024gemini} and claude-3-sonnet \cite{anthropic2024claude}) and two open-source LLMs (i.e., DeepSeek-V2-chat \cite{deepseekv2} and qwen1.5-72b-chat \cite{bai2023qwen}).

The results are shown in Table~\ref{tab:backbone}. Base refers to the version of AdaParser without both self-correction and SG-ICL.
We can observe that AdaParser consistently enhances the performance of the employed LLMs across all metrics by a large margin. On average, all LLMs improve by 5.1\%, 7.7\%, 19.9\%, and 23.6\% on GA, FGA, PA, and FTA, respectively. 
Meanwhile, LLMs with relatively weak performance benefit more from AdaParser, suggesting that AdaParser effectively compensates for deficiencies in backbone models.
These results demonstrate that AdaParser can be generally applied to different LLMs, maintaining high accuracy.

\begin{tcolorbox}
\textbf{Answer to RQ4:} AdaParser consistently improves LLMs’ performance of log parsing, even when utilizing relatively smaller LLMs, which demonstrates its generalizability.
\end{tcolorbox}

\subsection{RQ5: How efficient is AdaParser?}
\begin{figure}
    \centering
    \includegraphics[width=0.95\columnwidth]{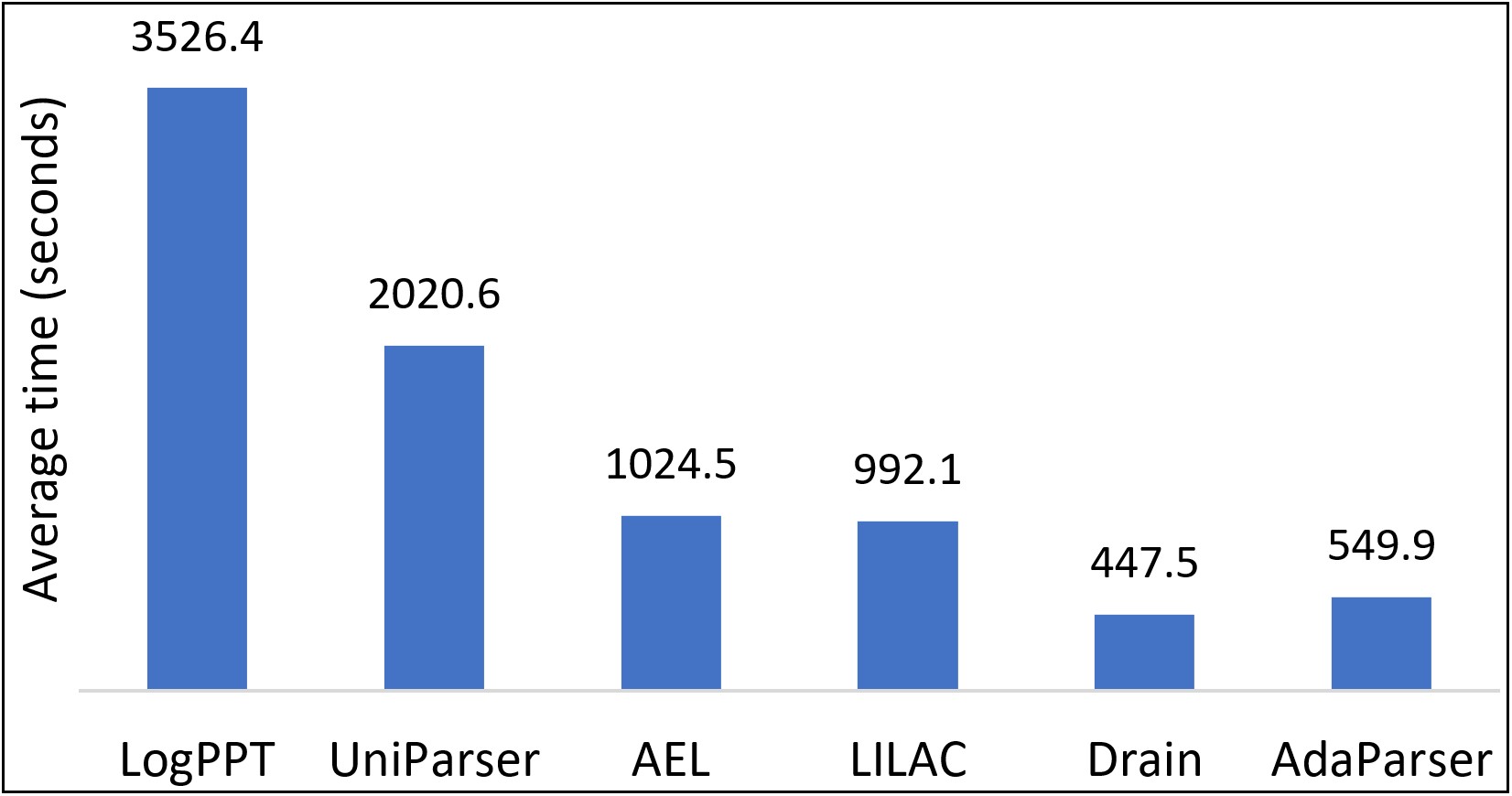}
    \caption{Efficiency of AdaParser and baselines on large-scale datasets.}
    \label{fig:time}
\end{figure}

Besides effectiveness, efficiency is another critical metric in the practical application of log parsers, given the substantial volume of logs \cite{wang2022spine,le2023alog,jiang2023large}. In this RQ, we assess the efficiency of AdaParser using the public large-scale Loghub-2.0 datasets as described in Sec. \ref{sec:datasets}.
To measure AdaParser's efficiency, we record the time it took to complete the parsing process for AdaParser and baselines on all log datasets. We then calculate the average parsing time for each parser, as depicted in \figurename~\ref{fig:time}.
 
The results indicate that the efficiency of AdaParser is on par with Drain, the most efficient syntax-based log parser currently available.
Specifically, AdaParser requires 549.9 seconds to process an average of 3.6 million logs, while this time of Drain is 447.5 seconds.
Conversely, semantic-based log parsers such as UniParser and LogPPT, despite the utilization of GPU acceleration, achieve low efficiency, trailing AdaParser by 3.67 and 6.41 times, respectively.

\begin{tcolorbox}
\textbf{Answer to RQ5:} AdaParser demonstrates superior efficiency in parsing large-scale logs, outperforming semantic-based log parsers by 3.67 to 6.41 times and achieving comparable performance to the fastest syntax-based log parsers.
\end{tcolorbox}

\section{Discussion}

\subsection{Applicability and Cost of AdaParser}

As demonstrated in Sec. \ref{sec:RQ4}, AdaParser is compatible with various LLMs, including open-source LLMs such as DeepSeek and Qwen. To address privacy concerns, we recommend deploying these open-source LLMs locally using vLLM \cite{kwon2023efficient}, a library designed for efficient LLM serving and inference. By simply modifying the \textit{api\_key} and \textit{base\_url} in AdaParser's configuration file, these LLMs can be seamlessly integrated into AdaParser. Therefore, we believe AdaParser is applicable in real-world scenarios.

AdaParser minimizes the cost of querying LLMs by employing a tree-based parser that reduces redundant queries. Additionally, for each query, since the number of tokens in a single log message is generally small (e.g., tens to hundreds of tokens), the cost of querying LLMs remains negligible.
Specifically, we calculate the token usage and cost of AdaParser on 14 benchmark datasets. The total input tokens are 1,221,754 and the total output tokens are 93,618. Given the pricing of \$0.5 per 1M input tokens and \$1.5 per 1M output tokens for gpt-3.5-turbo-0125 \cite{Pricing}, the total cost for parsing 3.6M log messages is only \$0.75. Note that the latest GPT-4o mini model, which is smarter and cheaper than GPT-3.5-Turbo, is priced at about one-third of GPT-3.5-Turbo. This means the cost of AdaParser can be further reduced.

\subsection{Threats to Validity}
We identified the following major threats to validity:

\textbf{Data Leakage.} Since LLMs are trained on extensive data, one potential threat is data leakage. Specifically, the LLM employed in AdaParser may have been trained on open-source log datasets, potentially leading to the memorization of ground-truth templates. However, our experiments indicate that AdaParser's performance in zero-shot scenarios is significantly inferior to that in few-shot scenarios, indicating a low probability of direct memorization. Additionally, AdaParser utilizes the gpt-turbo-3.5-0125 model for most experiments, which ceased updates before the release of Loghub-2.0 datasets. Thus, the probability of data leakage is negligible.

\textbf{Privacy Issue.} From the perspective of enterprises, logs are sensitive data because they typically contain extensive amounts of customer and operation information. Utilizing external LLMs to process internal logs can introduce significant privacy and security risks. 
As detailed in Sec.~\ref{sec:RQ4}, AdaParser is a general framework that supports the integration of various LLMs. Users can integrate their own LLMs into AdaParser, thereby avoiding privacy issues.

\textbf{Randomness.} The inherent randomness in LLM outputs is another potential threat to validity. This randomness manifests in (1) the initial templates generated by the SG-ICL-based parser and (2) the subsequent templates corrected by the template corrector.
To mitigate the former threat, we set the temperature and seed of LLMs to 0 to generate deterministic outputs for the same input. To mitigate the latter threat, we ran each experimental setting five times and used the mean of the results as the final results.

\section{related work}
Existing Log parsers can be categorized into syntax-based and semantic-based log parsers.
Syntax-based log parsers leverage human-crafted features to extract log templates, which have been widely explored in the past. Specifically, syntax-based log parsers can be further subdivided into three categories.
(1) \textit{Frequency-based} parsers: These log parsers \cite{vaarandi2003data,nagappan2010abstracting,vaarandi2015logcluster,dai2020logram} utilize frequent patterns of token position or n-gram information to distinguish templates and variables in log messages.
(2) \textit{Clustering-based} parsers: These log parsers \cite{fu2009execution,tang2011logsig,shima2016length,hamooni2016logmine} compute similarities between log messages to cluster them into different groups and then extract the constant parts of log messages.
(3) \textit{Heuristic-based} parsers: These log parsers \cite{du2016spell,he2017drain,jiang2008abstracting,makanju2009clustering,yu2023brain,liu2024xdrain} encode domain knowledge into general and effective heuristic rules to identify log templates. 
Semantic-based log parsers are a recently emerging class of parsers and can achieve higher parsing accuracy by mining semantics from log messages. These log parsers \cite{liu2022uniparser,huo2023semparser,li2023did,le2023alog,yu2023self,yu2023log,ma2024llmparser,wu2024logptr} typically need labeled log data for model training or tuning. For example, Uniparser \cite{liu2022uniparser} formulates log parsing as a token classification problem, employing Bi-LSTM models for training.
VALB \cite{li2023did} formulates log parsing as a sequence tagging problem and uses Bi-LSTM-CRF models for training.
LogPTR \cite{wu2024logptr} formulates log parsing as a text summarization problem and uses a pointer mechanism to copy words from log messages.
LogPPT \cite{le2023alog} tunes a pre-trained language model (e.g., RoBERTa) to perform log parsing.

However, recent studies \cite{khan2022guidelines,jiang2023large,petrescu2023log} have shown that existing log parsers struggle with large-scale and complex log data. Given the strong capacities of Large language models (LLMs) in understanding natural language and solving diverse tasks \cite{zhao2023survey}, several studies \cite{le2023blog,jiang2023lilac,liu2023logprompt,xu2024divlog} have explored the use of LLMs to improve parsing accuracy.
Le et al. \cite{le2023blog} were pioneers in evaluating the performance of LLMs in log parsing, which demonstrates their potential in performing log parsing. Xu et al. \cite{xu2024divlog} proposed DivLog, a method that utilizes the ICL capability of LLMs to enhance log parsing effectiveness. Jiang et al. \cite{jiang2023lilac} further combined the LLM-based parser with an adaptive parsing cache, achieving accurate and efficient LLM-based log parsing. 
Despite these advancements, existing LLM-based parsers rely solely on the responses generated by LLMs as the final parsed results. This can lead to inaccuracies due to the generation of plausible but incorrect templates or overly broad templates, thus reducing parsing accuracy.
Moreover, these methods require selecting demonstrations from a fixed candidate set sampled from extensive historical logs. However, in practice, frequent software updates often result in evolving logs, making sufficient historical logs unavailable. In contrast, our method AdaParser addresses these limitations by leveraging the proposed SG-ICL and self-correction mechanism, which significantly enhances both the accuracy and adaptability of LLM-based log parsing.

\section{conclusion}
This paper proposes AdaParser, an effective and adaptive log parsing framework using LLMs with SG-ICL and self-correction.
To facilitate accurate log parsing, AdaParser employs the LLM to self-correct potential parsing errors in its generated templates. Furthermore, AdaParser maintains a dynamic candidate set composed of previously generated templates as demonstrations to adapt evolving log data.
Extensive experiments on public large-scale datasets show that AdaParser outperforms state-of-the-art methods, even in zero-shot scenarios. When integrated with different LLMs, AdaParser consistently enhances the performance of the utilized LLMs by a large margin.
We believe AdaParser would benefit both practitioners and researchers in the field of log analysis.


\bibliographystyle{IEEEtran}
\bibliography{mybibfile}

\end{document}